\documentstyle[preprint,aps,epsf,eqsecnum]{revtex}
\begin{document}

\tightenlines

\def\lqcd{\Lambda_{\rm QCD}}
\def\xslash#1{{\rlap{$#1$}/}}
\def\dsl{\,\raise.15ex\hbox{/}\mkern-13.5mu D}
\preprint{\vbox{\hbox{UTPT-- 97-19}
\hbox{hep-ph/9710513}}}

\title{Corrections to Moments of the Photon Spectrum in the Inclusive Decay
  $B\to X_s \gamma$}

\author{Christian Bauer}

\address{\medskip Department of Physics, University of Toronto\\60
  St.~George Street, Toronto, Ontario,
Canada M5S 1A7 \medskip}

\bigskip
\date{October 1997}

\maketitle

\begin{abstract} We investigate the two main uncertainties on the
  extraction of the
  nonperturbative matrix elements $\bar{\Lambda}$ and $\lambda_1$ from
  moments of the  
  inclusive rare FCNC decay $B \to X_s \gamma$. The first one is due
  to unknown matrix elements of higher dimensional operators which are
  estimated by varying the matrix elements in a range as suggested by
  dimensional analysis. The effect of these terms is found to be
  small. A second uncertainty arises from a cut on the photon energy
  and we give model independent
  bounds on these uncertainties as well as their estimates from a
  simplified version of the ACCMM model. 
\end{abstract}

\pagebreak

\section{Introduction}
Heavy Quark Effective Theory (HQET) \cite{hqet1,hqet2} has proven to
be a useful tool to calculate inclusive decays of hadrons containing
one heavy quark \cite{inclusive,chay}. The presence of two widely separated
scales, the mass of the
heavy quark, $m_Q$, and $\lqcd$, allows us to perform
an operator product expansion (OPE) on the inclusive decay rates. At leading
order this reproduces
the parton model result with higher order terms suppressed by
powers of $\lqcd/m_Q$. To order
$(\lqcd/m_Q)^2$ they are parametrized by 
\begin{eqnarray}\label{dimension5}
\lambda_1 &=& \frac{1}{2 M} \langle H(v) | \bar{h}_v (iD)^2 h_v | H(v)
\rangle \nonumber\\
\lambda_2 &=& \frac{1}{2 d_H M} \langle H(v) | \bar{h}_v
  \frac{1}{2} (-i \sigma_{\mu\nu})  G^{\mu\nu} h_v | H(v) \rangle,
\end{eqnarray}
and 
\begin{equation}\label{lambdabar}
\bar{\Lambda} = M - m_Q + \frac{\lambda_1 + d_H \lambda_2}{2m_Q} +
{\cal O} \left(\frac{1}{m_Q^2}\right).
\end{equation}
Here $h_v$ is the spinor of a heavy quark with velocity $v$ and pole
mass $m_Q$,
$M$ is the mass of the heavy meson and $d_H=3\:(-1)$ for pseudoscalar
(vector) mesons. The gluon field strength is defined by $G_{\mu\nu} =
\left[iD_\mu,iD_\nu\right]$ and the covariant derivative by $D_\mu =
\partial_\mu - i g T^a\! A^a_\mu$. Knowledge of the
values of these parameters is important for a reliable
extraction of the CKM matrix elements $V_{cb}$ and
$V_{ub}$. The matrix element $\lambda_2$ can be extracted from the
observed mass splitting between the vector and the pseudo-scalar mesons,
\begin{equation}
\lambda_2=\frac{M_{B^*}^2-M_B^2}{4}\approx 0.12 \, {\rm GeV}^2,
\end{equation}
whereas no such simple relation exists for $\lambda_1$ and
$\bar{\Lambda}$. 

Many attempts have been made to extract values of these
parameters. Two model independent methods have been suggested using
moments of the electron energy spectrum \cite{electspect} and the invariant
hadronic mass spectrum \cite{hadmass} in the inclusive decay $B \to
X_c e \bar{\nu}$. In both approaches the moments are proportional to
linear combinations
of the two unknown matrix elements. Unfortunately, for the electron
energy spectrum, the two moments are given by almost the same
linear combination of $\bar{\Lambda}$ and $\lambda_1$, thus, even
small corrections to the moments lead to large uncertainties on the
parameters \cite{mcubed}. For the invariant mass spectrum, the two
moments determine two
almost orthogonal linear combinations, but corrections
suppressed by $(\Lambda_{QCD}/m_b)^3$ are huge for the second
moment. This yields
uncertainties for $\bar{\Lambda}$ and $\lambda_1$ comparable to those
from the electron energy spectrum \cite{hadcubed}. 

In Ref. \cite{ligeti} it has been shown that the mean photon energy
and its variance in the inclusive decay $B \to X_s \gamma$, which are
given to order $1/m_b^2$ by 
\begin{eqnarray}\label{moments}
\langle E_\gamma\rangle&=&\frac{m_b}{2}\left(1-\frac{ \lambda_1+
      3\lambda_2}{2m_b^2}\right)=\frac{M_B-\bar{\Lambda}}{2},
  \nonumber\\
  {\rm var}(E_\gamma) &=&-\frac{ \lambda_1}{12},
\end{eqnarray}
where ${\rm var}(E_\gamma)=\langle E_\gamma^2 \rangle - \langle E_\gamma
\rangle ^2$, determine the two parameters independently, thus, this analysis yields
two orthogonal bands in the $\bar{\Lambda}$--$\lambda_1$
plane. 
In this paper we study the theoretical uncertainties on the extraction
of $\lambda_1$ and $\bar{\Lambda}$ from this decay. Besides
perturbative corrections, which we do not consider in this paper,
there are higher order non--perturbative
corrections and we estimate their effects by calculating the $1/m_b^3$
corrections to the first two moments. The unknown matrix elements of the
dimension seven operators are varied
in their expected range $\pm \lqcd^3$. Constraints on some
of the matrix elements can be obtained from the known mass splitting
of vector and pseudoscalar mesons and from a vacuum saturation
approximation. They will be discussed in the following
sections. In addition, the measurement of the
differential decay spectrum involves imposing a cut on the photon
energy close to its endpoint \cite{kincl}.  As the OPE breaks down
close to the endpoint of the
photon spectrum, we anticipate errors which depend on this
cut. We will give model independent bounds on those
uncertainties as
well as estimates using the ACCMM model \cite{ACCMM}. 

\section{The operator product expansion and matrix elements}
The differential decay rate of the inclusive decay $B \to X_s \gamma$
is given by 
\begin{equation}
d\Gamma  = \frac{1}{2M_B} d[P.S.]_{pert} \langle 0 | R^\dagger |
\gamma \rangle \langle \gamma | R |0\rangle W(q),
\end{equation}
where 
\begin{equation}
W(q)=\sum_{X_s} d[P.S.]_{hadr} (2\pi)^4 \delta^4(P_B-P_X-q) \langle B |
  {\cal O}^\dagger |X_s \rangle \langle X_s | {\cal O} | B \rangle.
\end{equation}
$R$ and ${\cal O}$ are the radiative and hadronic part of the operator
that mediates the
decay, respectively. The matrix element of $R$ 
can be calculated perturbatively whereas $W(q)$ is related 
to the forward scattering amplitude 
\begin{equation}
T(q)=\langle B | T \{{\cal O}^\dagger {\cal O} \} | B \rangle
\end{equation}
via the optical
theorem
\begin{equation}
W(q)=2\,\rm{Im}T(q).
\end{equation}
As has been argued in \cite{inclusive,chay}, the time--ordered product can be
expanded by
performing an OPE 
\begin{equation}
T \left\{ {\cal O}^\dagger {\cal O} \right\} = \frac{1}{m_b} \left[
  {\cal O}_0 + \frac{1}{2 m_b} {\cal O}_1 + \frac{1}{4 m_b^2} {\cal
  O}_2 + \ldots \right].
\end{equation}
The first three terms in this expansion have been calculated for an
arbitrary 
hadronic operator of the form ${\cal O} = \bar{b} \Gamma s$ in Ref. 
\cite{rare_mike}. We
have extended this result by calculating the $1/m_b^3$ contribution to
the time--ordered product, ${\cal O}_3$. The general expressions for
these operators are quite lengthy and we present them in Appendix
\ref{app}. 

To obtain $T(q)$, we have to
determine matrix elements of these operators. Following
Ref. \cite{rare_mike} closely, the matrix elements at leading order,
\begin{equation}
  \langle B | \bar{b} \Gamma b | B \rangle,
\end{equation}
are only nonzero for $\Gamma=1$ or
$\Gamma=\gamma^\mu$. The matrix element of the conserved vector
current is given by 
\begin{equation}
  \langle B | \bar{b} \gamma^\mu b | B \rangle = 2 P_B^\mu,
\end{equation}
where $P_B^\mu=M_B v^\mu$ is the momentum of the heavy $B$ meson. Because
of this relation the dimension three operators are most conveniently
written in terms of the full QCD spinors $b$, rather than the effective
spinors $h_v$. At higher orders in the
$1/m_b$ expansion the spinors are expanded and written in
terms of $h_v$. Because of Lorentz invariance, the matrix elements
of the dimension 4 operators can be written as
\begin{equation}
  \langle B(v) | \bar{h}_v \Gamma D_\mu h_v | B(v) \rangle = v_\mu
  \langle B(v) | \bar{h}_v \Gamma (iv \cdot D) h_v | B(v) \rangle, 
\end{equation}
and vanish at leading order due to the equation of motion of the heavy
quark field 
\begin{equation}
  (iv \cdot D) h_v =0.
\end{equation}
At higher orders in the $1/m_b$ expansion corrections to this arise
which can most easily be 
incorporated by using the equation of motion to order $1/m_b$ 
\begin{equation}
  \bar{h}_v (iv \cdot D) h_v = -\frac{1}{2m_b} \bar{h}_v (iD)^2 h_v
+ \frac{1}{4m_b} \bar{h}_v
   i \sigma_{\mu\nu} G^{\mu\nu} h_v + {\cal O}\left(\frac{1}{m_b^2}\right).
\end{equation}
Matrix elements of general dimension five
operators can be parameterized
by
\begin{equation}
  \langle B(v) | \bar{h}_v \Gamma iD_\mu iD_\nu h_v | B(v) \rangle = M_B {\rm
  Tr} \left\{\Gamma P_+ \left(\frac{1}{3} \lambda_1 (g_{\mu\nu} - v_\mu v_\nu)
  +\frac{1}{2} \lambda_2 i \sigma_{\mu\nu}\right) P_+\right\},
\end{equation}
where $P_+=\frac{1}{2}(1+\rlap/v)$ is the projector which projects
onto the effective spinor $h_v$. Finally, the dimension six operators can
be parameterized by the matrix elements of two local operators
\cite{mcubed,mannel} 
\begin{eqnarray}\label{dimension6}
\frac{1}{2M_B}\langle B(v)| \bar{h}_v (iD_\alpha) (iD_\mu) (iD_\beta)
h_v | B(v)\rangle&=&\frac{1}{3}\rho_1\left(g_{\alpha\beta}-v_\alpha
v_\beta\right) v_\mu, \nonumber\\
 \frac{1}{2M_B}\langle B(v)| \bar{h}_v
(iD_\alpha) (iD_\mu) (iD_\beta) \gamma_\delta \gamma_5 h_v
| B(v)\rangle&=&\frac{1}{2} \rho_2 i\epsilon_{\nu\alpha\beta\delta}
v^\nu v_\mu 
\end{eqnarray}
and by matrix elements of two time--ordered products
\begin{eqnarray}\label{timeordered}
\frac{1}{2M_B} \langle B(v)|\bar{h}_v (iD)^2h_vi\int d^3x\int_{-\infty}^0 \!\!\!\!{\cal
L}_I(x)| B(v)\rangle+h.c.&=&\frac{{\cal T}_1 + 3 {\cal
T}_2}{m_b}, \nonumber\\ 
\frac{1}{2M_B} \langle B(v)|\bar{h}_v
\frac{1}{2}(-i \sigma_{\mu\nu})G^{\mu\nu} h_vi\int d^3x\int_{-\infty}^0
  \!\!\!\!{\cal L}_I(x)| B(v)\rangle+h.c.&=&\frac{{\cal T}_3+3{\cal
T}_4}{m_b}.
\end{eqnarray}

\section{Higher order corrections to the differential decay rate}
\begin{figure}
\centerline{\epsfxsize=10 cm \epsfbox{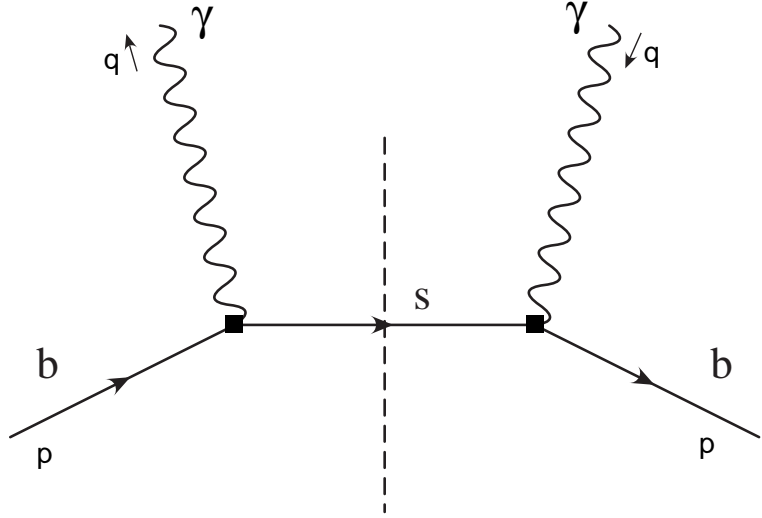}}
\caption{The Feynman diagram for the forward scattering amplitude.}
\label{forward}
\end{figure}

Integrating out the $W$ boson and the top quark \cite{rare_eff}, the
FCNC transition $b \to s$ is governed by the effective Hamiltonian
\begin{equation}\label{eff_hamilton}
{\cal H}_{eff} = -\frac{4 G_F}{\sqrt{2}} V_{tb}V_{ts}^* \sum_j C_j(\mu)
{\cal O}_j(\mu).
\end{equation}
In the leading logarithmic expansion, the
decay $B \to X_s \gamma$ occurs only through the operator 
\begin{eqnarray}\label{O_7}
{\cal O}_7&=&\frac{e}{16 \pi^2}m_b \bar{s}_{L_\alpha} \sigma^{\mu\nu}
b_{R_\alpha} F_{\mu\nu},
\end{eqnarray}
which leads at the parton level to the total decay rate 
\begin{equation}\label{partonrate}
\Gamma_0 = \frac{G_F^2 \left| V_{tb} V_{ts}^* \right|^2 \alpha\,
  C_7 (\mu)^2}{32 \pi^4} m_b^5.
\end{equation}
Higher order terms in the OPE arise from the expansion of the s--quark
propagator in Fig.\ref{forward} 
\begin{equation}
\frac{1}{m_b \xslash{v} + \xslash{k} - \xslash{q} - m_s} =
\frac{\xslash{v} - \xslash{\hat{q}}+\hat{m}_s}{m_b \Delta} \sum_n
(-1)^n \left( \frac{(\xslash{v} - \xslash{\hat{q}}+\hat{m}_s) \xslash{k}} {m_b
  \Delta} \right)^n,
\end{equation}
where $\Delta = 1 - 2 v \cdot \hat{q} - \hat{m}_s^2$, as well as from
the expansion of the full spinors $b$
\begin{equation}
b = \left(1 + \frac{i \xslash{k}}{2m_b} - \frac{k^2}{8m_b^2} + \cdots
\right) h_v.
\end{equation}
Starting at order $1/m_b^3$, additional contributions arise from the
difference of the $B$ meson states in the full and the effective
theory \cite{mcubed}. At order $1/m_b^3$ they are parameterized by
\begin{eqnarray}\label{states}
\frac{1}{2 M_B} \left\langle B(v) \left| \bar{h}_v (iD)^2 h_v \right|
  B(v) \right\rangle &=&
\lambda_1 + \frac{{\cal T}_1 + 3 {\cal T}_3}{m_b}\nonumber\\
\frac{1}{6 M_B} \left\langle B(v) \left| \bar{h}_v \frac{1}{2} (-i
    \sigma_{\mu\nu}) G^{\mu\nu}
  h_v \right| B(v) \right\rangle &=& \lambda_2 + \frac{{\cal T}_3 + 3 {\cal
    T}_4}{3m_b},
\end{eqnarray}
where the matrix elements ${\cal T}_1$--${\cal T}_4$ have been defined
in (\ref{timeordered}). 

To calculate $T(q)$ for the operator ${\cal O}_7$, we now use the
expressions for the generic operators given in Appendix \ref{app}, with
$\Gamma_1 = P_L \sigma_{\mu\nu}$ and $\Gamma_2 = \sigma_{\rho\sigma}
P_R$ contracted with
$q^\nu q^\sigma$, the $q$ dependence of the matrix element of
$R$. Parameterizing
the matrix elements as outlined in the previous section and making the
replacements
\begin{eqnarray}
\lambda_1 &\to& \lambda_1 + \frac{{\cal T}_1 + 3 {\cal
    T}_3}{m_b}\nonumber\\
\lambda_2 &\to& \lambda_2 + \frac{{\cal T}_3 + 3 {\cal
    T}_4}{3m_b}
\end{eqnarray}
to incorporate (\ref{states}) we find
\begin{eqnarray}\label{T(q)}
T(q) &=& -16 M_B m_b (v \cdot \hat{q})^2 \left[
  \frac{1}{\Delta} - \frac{\lambda_1}{2m_b^2} \left(\frac{5 - 6 v
  \cdot \hat{q}}{3 \Delta^3} \right) + \frac{\lambda_2}{2m_b^2}
  \frac{3}{\Delta^2} \right.\nonumber\\
&& + \frac{\rho_1}{6 m_b^3} \left(\frac{20 (v \cdot
\hat{q})^2 -24 v \cdot \hat{q} +9}{\Delta^4}\right)  + \frac{\rho_2}{2 
m_b^3} \left( \frac{2 v \cdot \hat{q} + 1}{\Delta^3}\right) \nonumber\\
&& - \left. \frac{{\cal T}_1 + 3 {\cal
T}_2}{2m_b^3} \left(\frac{5 - 6 v
  \cdot \hat{q}}{3 \Delta^3} \right) + \frac{{\cal T}_3 + 3 {\cal T}_4}{2m_b^3}
  \frac{1}{\Delta^2}\right],
\end{eqnarray}
leading to the total decay rate 
\begin{equation}
\Gamma = \Gamma_0 \left( 1 + \frac{\lambda_1 - 9 \lambda_2}{2m_b^2} -
\frac{11\rho_1 - 27 \rho_2}{6 m_b^3} + \frac{{\cal T}_1 + 3 {\cal T}_2 -
3 \left({\cal T}_3 + 3 {\cal T}_4 \right)}{2m_b^3}\right).
\end{equation}
Here $\Gamma_0$ is the parton level decay rate as defined in
(\ref{partonrate}). To order
$1/m_b^2$ this is in agreement with the expressions obtained in
Ref. \cite{rare_mike,early_b_s}.  

From the expression of $T(q)$ (\ref{T(q)}) we can also obtain the
differential decay rate
\begin{equation}\label{diffdecay}
\frac{d\Gamma}{dx} = \Gamma \left[ \delta(1-x) + A_1 \delta'(1-x)
  + A_2 \delta''(1-x) + A_3 \delta'''(1-x)\right],
\end{equation}
where
\begin{eqnarray}\label{Aterms}
A_1 &=& - \frac{\lambda_1 + 3 \lambda_2}{2m_b^2} - \frac{5\rho_1 - 21
    \rho_2}{6m_b^3} - \frac{{\cal T}_1 + 3 {\cal T}_2 + {\cal T}_3 + 3
    {\cal T}_4}{2m_b^3} \nonumber\\
A_2 &=& -\frac{\lambda_1}{6m_b^2} - \frac{2 \rho_1 - 3 \rho_2}{6m_b^3}
    - \frac{{\cal T}_1 + 3 {\cal T}_2}{6 m_b^3} \nonumber\\
A_3 &=& -\frac{\rho_1}{18m_b^3},
\end{eqnarray}
and $x = 2 E_\gamma / m_b$.
Again, to order $1/m_b^2$, this is in agreement with the expression
found in Ref. \cite{rare_mike}. We note that none of the coefficients of
the dimension  seven operators is
anomalously large compared to the ones of the dimension six
operators. This indicates that the $1/m_b^3$ corrections give rise to
only small uncertainties on the moments of the photon energy
spectrum. This will be investigated in much more detail in the next
section. 
 
Recently, Voloshin has shown the existence of a correction to the total
decay rate arising from a local operator suppressed
by powers of
$\lqcd^2/m_c^2$, which is given by \cite{voloshin}
\begin{equation}
\Delta \Gamma = -\Gamma_0 \frac{\lambda_2 C_2}{9 m_c^2 C_7}.
\end{equation}
$C_7$ and $C_2$ are the coefficient of the operators $O_7$
defined in (\ref{O_7}) and 
\begin{equation}
{\cal O}_2 = \bar{s}_{L_\alpha} \gamma^\mu b_{L_\beta}
\bar{c}_{L_\beta} \gamma_\mu c_{L_\alpha},
\end{equation}
respectively. At the scale $\mu = m_b$ the values of these
coefficients are $C_2 = 1.11$, $C_7 = -0.32$ and the correction to the
total rate is only about 2.5\%, of the same order as the
corrections due to the leading $1/m_b$ corrections. Since the leading
terms in the expressions for the mean
photon energy and its variance are already suppressed by powers of
$1/m_b$, this operator could have a significant effect on their
values. Fortunately, the terms $A_1$--$A_3$, which contribute to the
higher moments of the spectrum, arise from multiple poles $(1/\Delta)^n$
and are therefore suppressed by additional powers of $1/m_b$. The
leading contribution due to this new operator are suppressed
at least by $1/(m_b m_c^2)$ and up to this order we find
\begin{eqnarray}\label{voloshincontr}
\Delta \Gamma &=& \Gamma_0 \left[- \frac{\lambda_2 C_2}{9 m_c^2 C_7} -
\frac{\rho_1 C_2}{27 m_b m_c^2 C_7} + \frac{13 \rho_2 C_2}{36 m_b
m_c^2 C_7} \right] \nonumber\\
\Delta A_1 &=& -\frac{\rho_2 C_2}{9 m_c^2 m_b C_7}\nonumber\\
\Delta A_2 &=& \Delta A_3 = 0.
\end{eqnarray}
These terms are of the same order as the contributions from dimension
seven operators (\ref{Aterms}) and will be included in the determination
of the uncertainties on the extraction of the parameters
$\bar{\Lambda}$ and $\lambda_1$. 

\section{Uncertainties on $\bar{\Lambda}$ and $\lambda_1$}
To order $1/m_b^3$ the mean photon energy and its variance can be
obtained from (\ref{diffdecay}), (\ref{Aterms}) and (\ref{voloshincontr})
\begin{eqnarray}\label{momentscubed}
\langle E_\gamma \rangle &=& \frac{M_B - \bar{\Lambda}}{2} - \frac{13
  \rho_1 - 33 \rho_2}{24 m_b^2} - \frac{{\cal T}_1 + 3 {\cal T}_2
  +{\cal T}_3 + 3 {\cal T}_4}{8m_b^2} - \frac{\rho_2 C_2}{18 m_c^2
  C_7} \nonumber\\
{\rm var}(E_\gamma) &=& -
  \frac{\lambda_1}{12} - \frac{2 \rho_1 - 3 \rho_2}{12 m_b} -
  \frac{{\cal T}_1 + 3 {\cal T}_2}{12m_b}.
\end{eqnarray}
The size of the matrix elements $\rho_1$, $\rho_2$ and ${\cal
  T}_1$--${\cal T}_4$ are unknown, but by dimensional analysis we
expect them to be of order $\pm \lqcd^3$. Based on a vacuum
saturation approximation $\rho_1$ is assumed to be positive. 

The relation (\ref{lambdabar}) for the mass splitting between the
meson mass $M$
and the
quark mass $m_Q$ also has to be extended to include the $1/m_Q^3$
corrections. To this order it is given by \cite{mcubed}
\begin{equation}
M = m_Q + \bar{\Lambda} - \frac{\lambda_1 + d_H \lambda_2}{2m_Q} +
\frac{\rho_1 + d_H \rho_2}{4m_Q^2} - \frac{{\cal T}_1 + {\cal T}_3
  +d_H \left({\cal T}_2 + {\cal T}_4 \right)}{4m_Q^2}.
\end{equation}
Using the $\Delta M_B=M_{B^*}-M_B$ and the $\Delta M_D=M_{D^*}-M_D$
mass splitting, the relation
\begin{equation}\label{constraint}
\rho_2 - {\cal T}_2 - {\cal T}_4 = \frac{\kappa(m_c)M_B^2 \Delta M_B (M_D +
\bar{\Lambda}) - M_D^2 \Delta M_D (M_B + \bar{\Lambda})}{M_B +
\bar{\Lambda} - \kappa(m_c) (M_D + \bar{\Lambda})},
\end{equation}
where $\kappa(m_c) = (\alpha(m_c)/\alpha(m_b))^{3/\beta_0}$, can be used
to eliminate one of the 6 unknown parameters. 

\begin{figure}
\centerline{\epsfxsize=15 cm \epsfbox{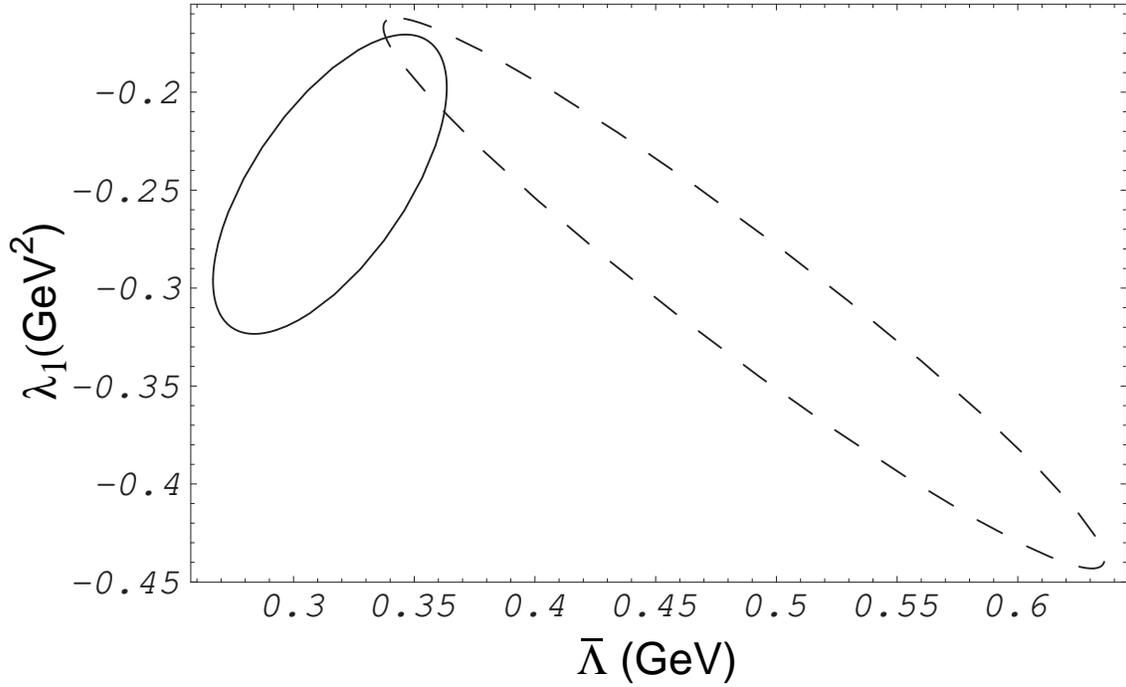}}
\caption{The estimated size of the uncertainties due to $1/m_b^3$
corrections on $\lambda_1$ and
$\bar{\Lambda}$ from $B \to X_s \gamma$ (solid) and from semileptonic
decay (dashed) using the method described in the text . The position
of the two ellipses has no meaning, only
the relative sizes.}
\label{compare}
\end{figure}

In order to compare the uncertainties on the values of $\bar{\Lambda}$
and $\lambda_1$ from dimension seven operators in this analysis with
the corresponding ones in the semileptonic decay $B \to X_c e
\bar{\nu}$, we estimate the effect of these higher dimensional terms
in the same way as in \cite{mcubed,hadcubed}. Since the current
measurement of the photon energy spectrum does not constrain the first two
moments to any reasonable accuracy, we use the hypothetical
data $\langle E_\gamma \rangle = 2.45 \, {\rm GeV}$ and ${\rm var}
(E_\gamma) = 0.13 \, {\rm GeV}^2$  in the following analysis (these are
obtained from the central 
values of $\lambda_1$ and $\bar{\Lambda}$, extracted from
inclusive semileptonic $B$ decays, by using Eq. (\ref{moments})). With
this input, we
now use Eq. (\ref{momentscubed}) to extract $\bar{\Lambda}$ and
$\lambda_1$ and 
determine how their values are affected by the dimension seven
operators by
randomly varying the
magnitude of the
parameters $\rho_1$, $\rho_2$ and ${\cal T}_1$--${\cal T}_4$ in
between
$0$ and $(0.5\,{\rm GeV})^3$ and imposing the positivity of $\rho_1$ as
well as the constraint (\ref{constraint}). In Fig.\ref{compare} we
show an ellipse in the $\bar{\Lambda}$ -- $\lambda_1$
plane which is centered about the mean of the distribution and which
contains 68\% of the points. This should give a reasonable estimate of
the theoretical uncertainty in the extraction of $\bar{\Lambda}$ and
$\lambda_1$ due to higher order corrections. As a comparison we also
show the corresponding ellipse obtained from the
hadron invariant mass  spectrum of the inclusive decay $B \to X_c e
\bar{\nu}$ \cite{mcubed}. Of course, the relative position of the
two ellipses has
no meaning since we have not used experimental values of the moments
for the photon spectrum. Only the relative size can be compared. The
standard deviation of the two parameters is given by
\begin{equation}
\Delta\bar{\Lambda}\approx 0.03 \,{\rm GeV}
\end{equation}
and 
\begin{equation}
\Delta\lambda_1\approx 0.05  \,{\rm GeV}^2.
\end{equation}
These uncertainties are much smaller than the expected size of the
matrix elements, thus, the
corrections should be well under control.

\section{Uncertainties from the photon energy cut}
A cut on the photon energy to suppress the strong background from
other $B$ decays reduces the amount of phase space considerably. In
its measurement of the inclusive photon energy spectrum \cite{kincl},
CLEO imposed a cut at 2.2 GeV, about 500 MeV below the maximum photon
energy. The OPE is only valid if smeared over a region much larger than
$\lqcd$, but it has been shown \cite{shape} that for a smearing
region of order $\lqcd$ the
most singular terms in the OPE, which are formally of the same order,
can be resummed in the so called shape function. For an even smaller
smearing region of order $\lqcd^2/m_b$ all the subleading terms become
equally important.

Since the OPE gets worse the closer the photon energy cut gets to the
endpoint of the spectrum, one expects uncertainties that depend on the
value of this cut. The expression for the differential decay rate
obtained by performing an OPE contains only delta functions and their
derivatives which contribute at $E_\gamma=m_b/2$. A photon energy cut
affecting the lower bound on the range of
integration does not change the results for the moments as long it is
below $E_\gamma=m_b/2$. The
effect of the cut is hidden in the fact that the OPE for the
differential decay rate gets worse as the smearing region
diminishes. The experimental value of the higher moments of the
photon energy spectrum measured in the presence of a cut will be
larger than the value which determines the parameters $\bar{\Lambda}$
and $\lambda_1$ as defined in (\ref{moments}). Therefore, a cut will
shift the values of the two parameters extracted in this way, leading
to uncertainties on their values.

In this section we will take two different
approaches to investigate the size of those uncertainties. The first
will use the shape function analysis
to yield model independent bounds on
the uncertainties and in the second we will estimate them using a simple
version of the ACCMM model \cite{ACCMM}. 

\subsection{Model independent bounds on the uncertainties}
The nth moment ($n\geq1$) with a cut on the photon energy $E_0$ is given by
\begin{eqnarray}\label{moment}
M_n^{E_0} &=& \frac{\int_{E_0}^{E_{max}} E^n
  \frac{d\Gamma}{dE}dE}{\Gamma^{E_0}} \nonumber\\
&=& \frac{\int_0^{E_{max}} \theta(E - E_0)E^n \frac{d\Gamma}{dE}dE}
  {\Gamma^{E_0}},
\end{eqnarray}
where 
\begin{equation}
\Gamma^{E_0} = \int_{E_0}^{E_{max}} \frac{d\Gamma}{dE}dE
\end{equation}
is the total decay rate with a cut.
The positive powers of $E$ in the integrand of the numerator of
Eq. (\ref{moment}) weight the higher energy part of the spectrum
more, therefore
\begin{equation}
M_n^{E_0} \geq M_n,
\end{equation}
where $M_n$ is the moment without a photon energy cut. 

\begin{figure}
\centerline{\epsfxsize=15 cm \epsfbox{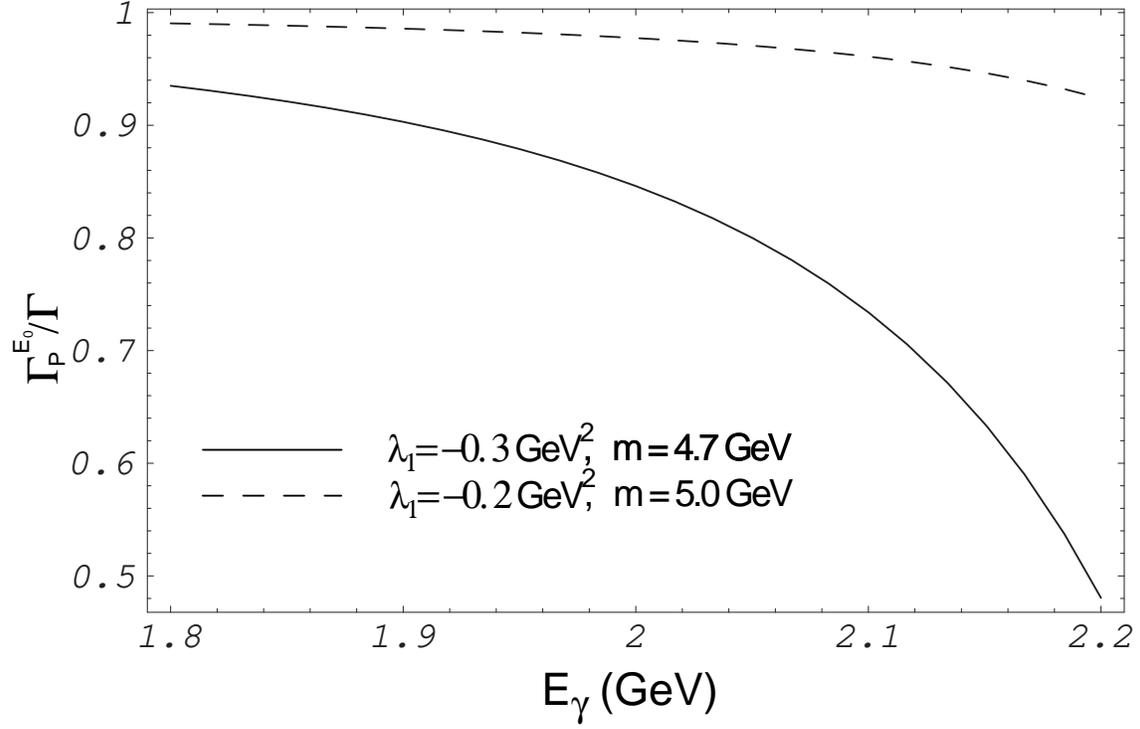}}
\caption{The model independent bound on the total decay rate.}
\label{totalbound}
\end{figure}

In \cite{vub} a method has been proposed to obtain model independent
bounds on the total decay rate. In this approach one replaces the step
function with a smooth function $P(E,E_0)$, obeying
\begin{eqnarray}\label{P_requirements}
P(E,E_0) \leq 1 \quad &{\rm for}& \quad E_0 \leq E \leq E_{max} \nonumber\\
P(E,E_0) \leq 0 \quad &{\rm for}& \quad 0 \leq E \leq E_0,
\end{eqnarray}
to find the inequality
\begin{equation}\label{inequality}
\Gamma^{E_0} \geq \Gamma^{E_0}_P = \int_0^{E_{max}} P(E,E_0)
\frac{d\Gamma}{dE} dE.
\end{equation}
The resulting bounds on the total decay rate are shown in
Fig.\ref{totalbound}. One can see that depending on the values of
$\lambda_1$ and $m_b$, a cut at 2.2 GeV can have a significant effect
on the measurement of the total decay rate. This is in agreement with
the analysis done using the ACCMM model \cite{alibound} which has been
used by the CLEO collaboration \cite{kincl}. For a monotonically
increasing function $P(E,E_0)$, this analysis can also be used to
obtain a bound on the moments as measured by experiment
\begin{equation}
\frac{\int_0^{E_{max}} P(E,E_0) E^n \frac{d\Gamma}{dE}
dE}{\Gamma^{E_0}_P} = M_{n_P}^{E_0} \geq
M_n^{E_0} \geq M_n.
\end{equation}
The most singular terms can be resummed into the shape function
\begin{equation}\label{shapefctn}
\frac{d\Gamma_s}{dE} = \Gamma \sum_{n=0}^\infty
\left(\frac{m_b}{2}\right)^n A_n \delta^{(n)}\left(\frac{m_b}{2}-E\right),
\end{equation}
where the coefficients $A_n$ now only contain the leading term in the
$1/m_b$ expansion. Using this we find
\begin{equation}\label{momentdef}
M_{n_P}^{E_0} = \frac{\sum_n \left(\frac{m_b}{2}\right)^n A_n
(E^n P(E,E_0))^{(n)} |_{E=m_b/2}}{\sum_n \left(\frac{m_b}{2}\right)^n A_n
(P(E,E_0))^{(n)} |_{E=m_b/2}}.
\end{equation}
If $P$ is a polynomial of order $k$, only the first $k$ terms
in the shape function (\ref{shapefctn}) contribute. 
In the following analysis we will use the polynomial 
\begin{equation}\label{Pdef}
P(E,E_0) = 1 - \left(\frac{E_1 - E}{E_1 - E_0}\right)^s \qquad E_0 < E_1.
\end{equation}
In order to have a monotonic function which satisfies the
requirements (\ref{P_requirements}), $E_0$ has to coincide with the
value of the photon energy cut and $E_1$ with the maximum photon
energy, $E_{max}$.

\begin{figure}
\centerline{\epsfxsize=15 cm \epsfbox{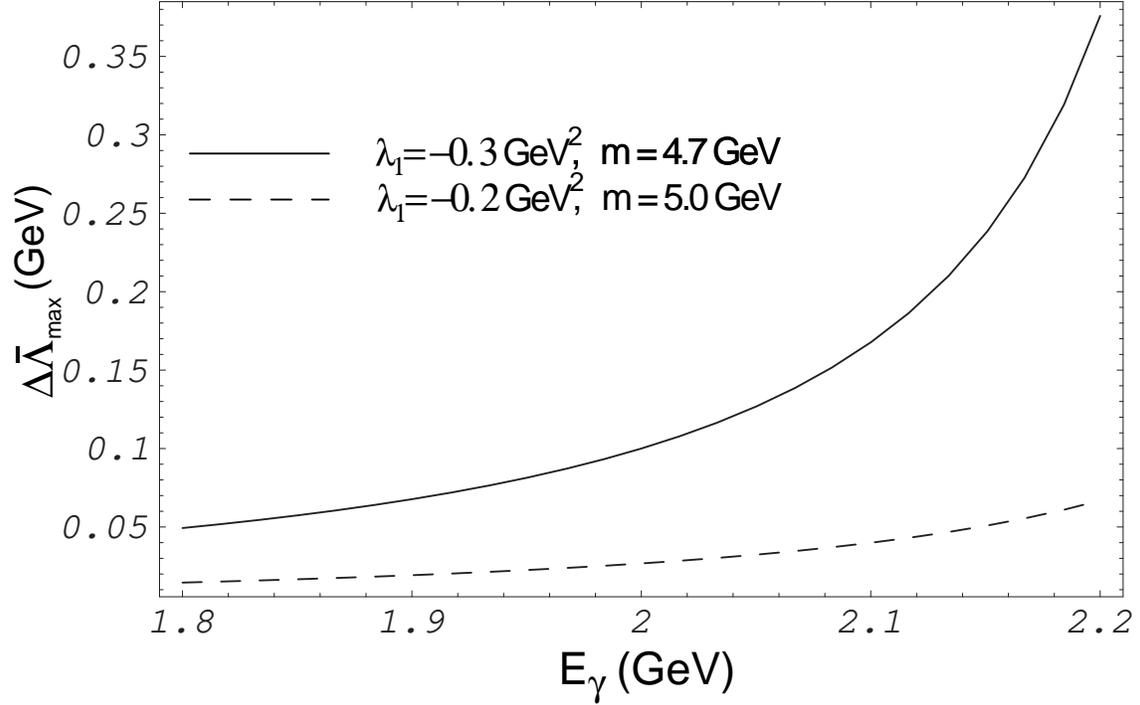}}
\caption{The model independent bound on the uncertainty of
$\bar{\Lambda}$.}
\label{firstbound}
\end{figure}

Since we are working to order $1/m_b^3$, we have to use a polynomial
of order $(3-n)$ to obtain a bound on the nth moment. For the first
moment we find using Eqns. (\ref{momentdef}) and (\ref{Pdef})
\begin{equation}
M_1^{E_0} \leq {\frac{3\,\left( 2\,E_0 - m_b \right) \,m_b\,
      \left( 2\,E_0 - 4\,E_m + m_b
        \right)  + \left( -4\,E_m + 3\,m_b \right)
        \,{{\lambda }_1} + {{\rho }_1}}{2\,
     \left( 3\,\left( 2\,E_0 - m_b \right) \,
        \left( 2\,E_0 - 4\,E_m + m_b
           \right)  + {{\lambda }_1} \right) }}
\end{equation}
From this result it is trivial to obtain a bound on the uncertainty on
$\bar{\Lambda}$. A plot of this bound  for $\rho_1=(0.3\,{\rm GeV})^3$
and two different sets of
values $\lambda_1$ and $m_b$ is presented in
Fig.\ref{firstbound}. For a cut at 2.2 GeV the bound on
the uncertainty on
$\bar{\Lambda}$ is between 100 MeV and 400 MeV, thus, it can be of the same order
as $\lqcd$, leading to large errors. In order to be
able to determine a precise value of
$\bar{\Lambda}$, the cut would have to be lowered to around 2
GeV. Here, the bound on the uncertainty is between 20 MeV and 100 MeV. 
Since $\lambda_1$ is related to the variance of the spectrum, which is
the difference of two moments, the P--function analysis can not be
used to bound its uncertainty. This is due to the fact that the
variance is sensitive to the tails of the spectrum. 

An inequality such as derived in (\ref{inequality}) holds true
only for a smooth, positive definite shape function. In general,
there is nothing that guarantees the validity of this assumption,
although models such as the ACCMM model \cite{ACCMM} predict a
positive shape function. It should also be pointed out that the inequality
(\ref{inequality}) is true rigorously only for smooth functions
$d\Gamma_s/dE$. In this approach we use a singular expansion of the
differential decay rate, so the convergence of this expansion is
essential. As said above, the shape function analysis is only
valid if the smearing region is of order $\lqcd$, a scale which also
sets the size of matrix elements of
higher dimensional operators. We therefore expect the bounds to break
down for values of the matrix elements which exceed the available
phase space considerably. The breakdown of the inequality
(\ref{inequality}) can be seen by using $E_m =
2.6$ GeV, $E_0 = 2$ GeV, $\lambda_1 = -0.2\,\rm{GeV}^2$ and $m_b =
4.8$ GeV. Inserting those values into (\ref{inequality}) we find 
\begin{equation}
0.193 \frac{\rho_1}{\rm{GeV}^3} + 0.917 =
\frac{\Gamma_P^{E_0}}{\Gamma} \leq 1
\end{equation}
which breaks down for $\rho_1 \geq (0.76\,\rm{GeV})^3$. 

Since we could not obtain a bound on the uncertainty on
$\lambda_1$ and we
are in no position to rigorously justify the assumptions we
made in this chapter we will now use a simple model to calculate the
effect of the photon energy cut on the first two moments. 

\subsection{Uncertainties using the ACCMM model}
We will use a simplified version of the ACCMM model \cite{ACCMM} to
estimate the
value of the moments as measured in the presence of a cut on the
photon energy. Assuming a Gaussian distribution of the relative
momentum of the b--quark inside the $B$ meson, neglecting the
mass of the s--quark and the momentum dependence of the
b--quark mass, the spectral
function is given by \cite{shape}
\begin{equation}
\frac{1}{\Gamma} \frac{d\Gamma}{dE} = \frac{1}{\sqrt{2\pi} \sigma_E}
\exp \left\{ - \frac{(E-\frac{m_b}{2})^2}{2 \sigma_E^2}\right\},
\end{equation}
where $\sigma_E^2 = -\frac{\lambda_1}{12}$. 

\begin{figure}
\centerline{\epsfxsize=15 cm \epsfbox{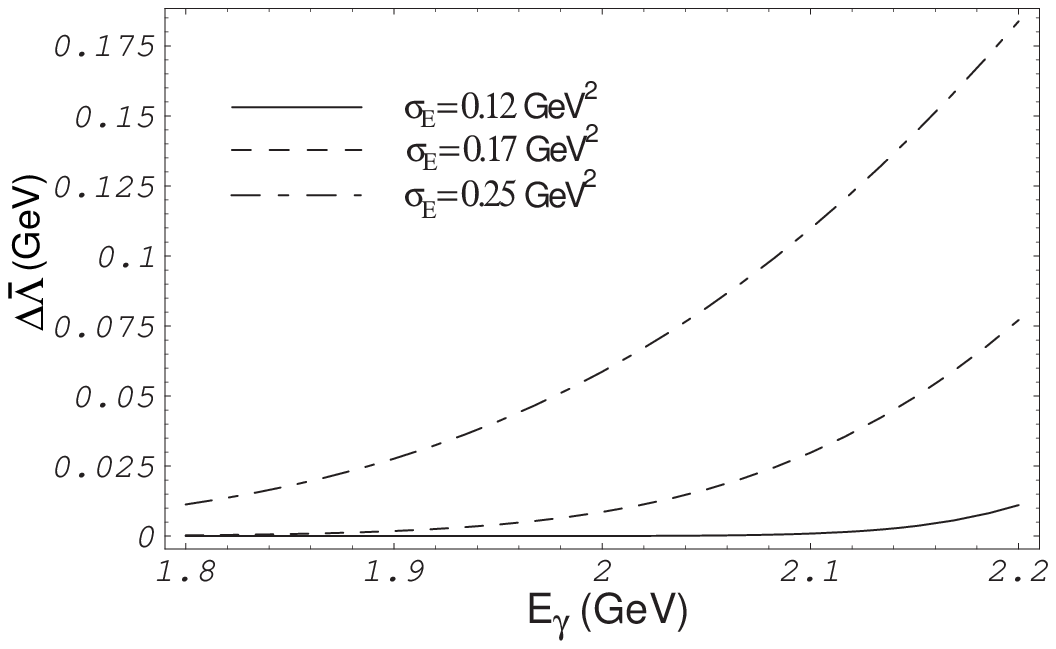}}
\caption{The uncertainty on $\bar{\Lambda}$ due to a photon energy cut from
  the ACCMM model.}
\label{firstacm}
\end{figure}

The difference between the first moment in the ACCMM model with and without a
cut gives us an estimate of the uncertainty on the value of
$\bar{\Lambda}$ and the result is shown in
Fig.\ref{firstacm} for different values of $\sigma_E$. For a cut at
2.2 GeV, the uncertainty on the
parameter $\bar{\Lambda}$ in this model is between 20 MeV and 180
MeV, depending on the width of the spectrum. Since the
real effect of the photon energy
cut could easily exceed this estimate by a
factor of two or three, a cut on the photon energy at
2.2 GeV could destroy the possibility of accurately determining the
value of $\bar{\Lambda}$. If the cut could be lowered to
$\approx$ 2 GeV, then an accurate extraction should be possible. This
is in agreement with the model independent results obtained in the
last section. 

\begin{figure}
\centerline{\epsfxsize=15 cm \epsfbox{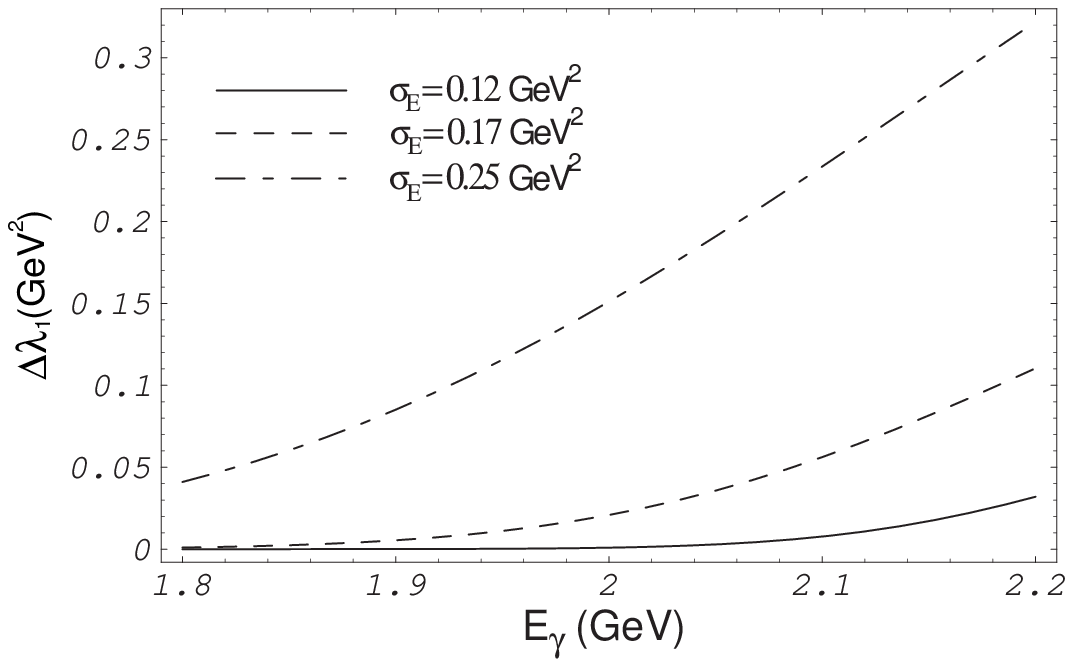}}
\caption{The uncertainty on $\lambda_1$ due to a photon energy cut from
  the ACCMM model.}
\label{secondacm}
\end{figure}

The result for a similar calculation for the variance of the spectrum
is shown in Fig.\ref{secondacm}. For a cut of 2.2 GeV, the uncertainty
on $\lambda_1$ in this model is between $0.05 \, {\rm GeV}^2$ and $0.3 \, {\rm
GeV}^2$, Again, considering the fact that this is only a model
calculation and it might underestimate the effect considerably, this
indicates that an extraction of $\lambda_1$
from the present CLEO measurement might be unreliable. Lowering the
cut to $\approx$ 1.9 GeV should enable a precise 
determination of this parameter from the decay $B \to X_s \gamma$.

\section{Conclusions}
We have analyzed the uncertainties on the extraction of the two
nonperturbative matrix elements $\bar{\Lambda}$ and $\lambda_1$ from
the mean photon energy and its variance in the inclusive decay \mbox{$B\to
X_s \gamma$}. 
Besides perturbative corrections which we have not
considered here, uncertainties arise from matrix elements of higher
dimensional operators which are suppressed by additional powers of
$\lqcd/m_b$. We have calculated the first two moments  up
to order $1/m_b^3$ and estimated the effect of the
unknown matrix elements by varying their values in their expected
range of magnitude.
The resulting uncertainties on $\bar{\Lambda}$ and $\lambda_1$ are
given by
\begin{equation}
\Delta\bar{\Lambda}\approx 0.03 \,{\rm GeV}
\end{equation}
and 
\begin{equation}
\Delta\lambda_1\approx 0.05  \,{\rm GeV}^2.
\end{equation}

A more serious uncertainty arises from the effect of a cut on the
photon
energy which has to be imposed in order to reduce the large background
from other processes. The differential decay rate as calculated via
the OPE is given in terms of a delta function and its derivatives
contributing at $E_\gamma = m_b/2$. The
effect of a cut on the photon energy therefore does not affect the
results for the
moments. It is hidden in the fact that the OPE breaks down as the cut
approaches the endpoint of the photon energy. We have used an approach
suggested in \cite{vub} to obtain a model independent bound on the
uncertainty on $\bar{\Lambda}$, whereas no such bound could be
derived for the uncertainty on $\lambda_1$. The bound
indicates that an accurate extraction of $\bar{\Lambda}$ is definitely
possible for a cut at 2 GeV, whereas for the present cut at 2.2 GeV
the errors might be large. We have also used a simplified version of
the ACCMM model to
estimate
the effect the photon energy cut. The uncertainties depend strongly on
the width of the spectrum in our model. We again find that an accurate
determination of $\bar{\Lambda}$ should be possible if the cut can be
lowered to about 2 GeV, Depending on the width of the spectrum, the
cut has to be lowered even further to allow a precise determination of
$\lambda_1$..

\acknowledgments
We are indebted to Michael Luke for numerous fruitful discussions on various
aspects of this analysis. We would also like to thank Mark Wise for
drawing our attention towards the investigation of the effects of the
photon energy cut. 

\appendix
\section{The general expressions for the first three operators in
the OPE}\label{app}

In this appendix we will present the first three terms in the
OPE for general operators
\begin{equation}
T\{\bar{b} \Gamma_1 s,\bar{s} \Gamma_2 b\} = \frac{1}{m_b} \left[{\cal
O}_0 + \frac{1}{2m_b} {\cal O}_1 + \frac{1}{4m_b^2} {\cal O}_2 +
\frac{1}{8m_b^3} {\cal O}_3 + \ldots \right].
\end{equation}
With the conventions
\begin{eqnarray}
D_\mu &=& \partial_\mu - i g T^a\! A^a_\mu \nonumber\\
G_{\mu\nu} &=& \left[iD_\mu,iD_\nu\right]
\end{eqnarray}
they are given by
\begin{eqnarray}
{\cal O}_0 &=& -\frac{1}{x} \bar{b} \Gamma_1 (\rlap/v - \rlap/\hat{q}
  +\hat{m}_s ) \Gamma_2 b\\
{\cal O}_1 &=& - \frac{2}{x} \bar{h} \Gamma_1 \gamma^\alpha \Gamma_2
  iD_\alpha h + \frac{4}{x^2} (v-\hat{q})^\alpha \bar{h} \Gamma_1
  (\xslash{v} - \hat{\xslash{q}} + \hat{m}_s) \Gamma_2 iD_\alpha h\\
{\cal O}_2 &=& -\frac{16}{x^3} (v-\hat{q})^\alpha (v-\hat{q})^\beta
  \bar{h} \Gamma_1 (\xslash{v} - \hat{\xslash{q}} + \hat{m}_s)
\Gamma_2 iD_\alpha iD_\beta h\nonumber\\
&& + \frac{4}{x^2} \bar{h} \Gamma_1 (\xslash{v} - \hat{\xslash{q}} + \hat{m}_s)
  \Gamma_2 (iD)^2 h\nonumber\\
&& + \frac{4}{x^2} (v-\hat{q})^\beta \bar{h} \Gamma_1 \gamma^\alpha
  \Gamma_2 (iD_\alpha iD_\beta + iD_\beta iD_\alpha) h\\
&& -\frac{2}{x^2} \hat{m}_s \bar{h} \Gamma_1 i \sigma_{\alpha\beta}
  \Gamma_2 G^{\alpha\beta} h\nonumber\\
&& - \frac{2}{x^2} i \varepsilon^{\mu\lambda\alpha\beta}
  (v-\hat{q})_\lambda \bar{h} \Gamma_1 \gamma_\mu \gamma_5 \Gamma_2
  G_{\alpha\beta} h\nonumber\\
&& - \frac{2}{x} \bar{h} (\gamma^\beta \Gamma_1 \gamma^\alpha \Gamma_2
  + \Gamma_1 \gamma^\beta \Gamma_2 \gamma^\alpha) iD_\beta iD_\alpha
  h\nonumber\\
&& + \frac{4}{x^2} (v-\hat{q})^\alpha \bar{h} \gamma^\beta \Gamma_1
  (\xslash{v} - \hat{\xslash{q}} + \hat{m}_s) \Gamma_2 iD_\beta iD_\alpha
  h\nonumber\\
&& + \frac{4}{x^2} (v-\hat{q})^\alpha \bar{h} \Gamma_1 (\xslash{v} -
  \hat{\xslash{q}} + \hat{m}_s) \Gamma_2 \gamma^\beta iD_\alpha iD_\beta h\\
{\cal O}_3 &=& 
\frac{8}{3x^2} \bar{h} \Gamma_1 \gamma^\alpha \Gamma_2
(iD_\alpha (iD)^2 + (iD)^2 iD_\alpha + iD_\beta iD_\alpha iD^\beta) h
\nonumber\\
&& - \frac{32}{3x^3} (v-\hat{q})^\rho (v-\hat{q})^\sigma \bar{h}
\Gamma_1 \gamma^\alpha \Gamma_2 (iD_\alpha iD_\rho iD_\sigma + iD_\rho
iD_\sigma iD_\alpha + iD_\rho iD_\alpha iD_\sigma) h \nonumber\\
&& - \frac{32}{3x^3} (v-\hat{q})^\alpha \bar{h} \Gamma_1 (\xslash{v} -
\hat{\xslash{q}} + \hat{m}_s) \Gamma_2 (iD_\alpha (iD)^2 + (iD)^2
iD_\alpha + iD_\beta iD_\alpha iD^\beta) h
\nonumber\\
&& + \frac{64}{3x^4} (v-\hat{q})^\alpha (v-\hat{q})^\rho
(v-\hat{q})^\sigma \bar{h} \Gamma_1 (\xslash{v} - \hat{\xslash{q}} +
\hat{m}_s) \Gamma_2 (iD_\alpha iD_\rho iD_\sigma + iD_\rho
iD_\sigma iD_\alpha + iD_\rho iD_\alpha iD_\sigma) h \nonumber\\
&& - \frac{16}{3x^3} (v-\hat{q})^\rho (v-\hat{q})^\alpha \bar{h}
\Gamma_1 \gamma^\beta \Gamma_2 (iD_\rho G_{\alpha\beta}) h \nonumber\\
&& - \frac{16}{3x^3} (v-\hat{q})^\beta \bar{h} \Gamma_1 (\xslash{v} -
\hat{\xslash{q}} + \hat{m}_s) \Gamma_2 (i D^\mu
G_{\mu\beta}) h \nonumber\\
&& + \frac{16}{3x^2} \bar{h} \Gamma_1 \gamma^\beta \Gamma_2 (i D^\mu
G_{\mu\beta}) h \nonumber\\
&& + \frac{2}{x^2} i \varepsilon^{\beta\rho\sigma\alpha} \bar{h}
\Gamma_1 \gamma_\rho \gamma^5 \Gamma_2 \{i D_\sigma,G_{\alpha\beta}\} h
\nonumber\\
&& + \frac{8}{x^3} \hat{m}_s(v-\hat{q})^\sigma \bar{h} \Gamma_1
i \sigma^{\alpha\beta} \Gamma_2 \{i D_\sigma , G_{\alpha\beta}\} h
\nonumber\\
&& + \frac{8}{x^3} i \varepsilon^{\beta\rho\alpha\sigma}
(v-\hat{q})_\sigma (v-\hat{q})_\lambda \bar{h} \Gamma_1 \gamma_\rho
\gamma^5 \Gamma_2 \{i D^\lambda , G_{\alpha\beta}\} h \nonumber\\
&& + \frac{1}{x} v^\alpha v^\beta \bar{h}
\Gamma_1 \gamma^\alpha \Gamma_2 (iD_\mu G_{\mu\beta}) h \nonumber\\
&& - \frac{1}{x} v^\alpha v^\beta \bar{h}
\Gamma_1 \gamma^\alpha \Gamma_2 i \sigma^{\rho\sigma} \{i D_\rho ,
G_{\sigma\beta}\} h \nonumber\\
&& - \frac{2}{x^2} (1-v \cdot q) v^\alpha \bar{h} \Gamma_1 (\xslash{v} -
\hat{\xslash{q}} + \hat{m}_s) \Gamma_2 (iD_\mu G_{\mu\alpha}) h
\nonumber\\
&& + \frac{2}{x^2} (1-v \cdot q) v^\alpha \bar{h} \Gamma_1 (\xslash{v} -
\hat{\xslash{q}} + \hat{m}_s) \Gamma_2 i \sigma^{\rho\sigma} \{i
D_\rho,G_{\sigma\alpha}\} h \nonumber\\
&& - \frac{1}{x} v^\alpha \bar{h} \Gamma_1 \gamma^\rho \Gamma_2
\gamma^\sigma  iD_\rho iD_\alpha iD_\sigma h \nonumber\\
&& -\frac{3}{x} \bar{h} \Gamma_1 \gamma^\alpha \Gamma_2 iD_\alpha
(iD)^2 h \nonumber\\
&& + \frac{3}{x} \bar{h} \Gamma_1 \gamma^\alpha \Gamma_2 i
\sigma^{\rho\sigma} iD_\alpha iD_\rho iD_\sigma h \nonumber\\
&& + \frac{2}{x^2} v^\alpha (v-\hat{q})^\beta \bar{h} \Gamma_1 (\xslash{v} -
\hat{\xslash{q}} + \hat{m}_s) \Gamma_2 \gamma^\rho iD_\beta iD_\alpha
iD_\rho h \nonumber\\
&& + \frac{6}{x^2} (v-\hat{q})^\alpha \bar{h} \Gamma_1 (\xslash{v} -
\hat{\xslash{q}} + \hat{m}_s) \Gamma_2 iD_\alpha (iD)^2 h \nonumber\\
&& - \frac{6}{x^2} (v-\hat{q})^\alpha \bar{h} \Gamma_1 (\xslash{v} -
\hat{\xslash{q}} + \hat{m}_s) \Gamma_2 i \sigma^{\rho\sigma} iD_\alpha iD_\rho
iD_\sigma h \nonumber\\
&& - \frac{1}{x} v^\alpha \bar{h} \gamma^\rho \Gamma_1 \gamma^\sigma \Gamma_2
iD_\rho iD_\alpha iD_\sigma h \nonumber\\
&& -\frac{3}{x} \bar{h} \Gamma_1 \gamma^\alpha \Gamma_2 (iD)^2 iD_\alpha
h \nonumber\\
&& + \frac{3}{x} \bar{h} i \sigma^{\alpha\beta} \Gamma_1 \gamma^\rho
\Gamma_2 iD_\alpha iD_\beta iD_\rho h \nonumber\\
&& + \frac{2}{x^2} v^\alpha (v-\hat{q})^\beta \bar{h} \gamma^\rho
\Gamma_1 (\xslash{v} - \hat{\xslash{q}} + \hat{m}_s) \Gamma_2
iD_\rho h iD_\alpha iD_\beta \nonumber\\
&& + \frac{6}{x^2} (v-\hat{q})^\alpha \bar{h} \Gamma_1 (\xslash{v} -
\hat{\xslash{q}} + \hat{m}_s) \Gamma_2 (iD)^2 iD_\alpha h \nonumber\\
&& - \frac{6}{x^2} (v-\hat{q})^\alpha \bar{h} i \sigma^{\rho\sigma}
\Gamma_1 (\xslash{v} - \hat{\xslash{q}} + \hat{m}_s) \Gamma_2 iD_\rho iD_\sigma
iD_\alpha h \nonumber\\
&& - \frac{2}{x} \bar{h} \gamma^\rho \Gamma_1 \gamma^\alpha \Gamma_2
\gamma^\beta iD_\rho iD_\alpha iD_\beta h \nonumber\\
&& + \frac{4}{x^2} (v-\hat{q})^\alpha \bar{h} \gamma^\rho \Gamma_1 (\xslash{v} -
\hat{\xslash{q}} + \hat{m}_s) \Gamma_2 \gamma^\beta iD_\rho iD_\alpha
iD_\beta h \nonumber\\
&& + \frac{4}{x^2} (v-\hat{q})^\beta \bar{h} \gamma^\rho \Gamma_1
\gamma^\alpha \Gamma_2 iD_\rho (iD_\alpha iD_\beta + iD_\beta
iD_\alpha) h \nonumber\\
&& + \frac{4}{x^2} \bar{h} \gamma^\rho \Gamma_1 (\xslash{v} -
\hat{\xslash{q}} + \hat{m}_s) \Gamma_2 iD_\rho (iD)^2 h \nonumber\\
&& - \frac{16}{x^3} (v-\hat{q})^\alpha (v-\hat{q})^\beta \bar{h}
\gamma^\rho \Gamma_1 (\xslash{v} -
\hat{\xslash{q}} + \hat{m}_s) \Gamma_2 iD_\rho iD_\alpha iD_\beta h
\nonumber\\
&& - \frac{2}{x^2} \hat{m}_s \bar{h} \gamma^\rho \Gamma_1 i
\sigma^{\alpha\beta} \Gamma_2 iD_\rho G_{\alpha \beta} h \nonumber\\
&& - \frac{2}{x^2} i \varepsilon^{\sigma\alpha\rho\beta}
(v-\hat{q})_\beta \bar{h} \gamma^\lambda \Gamma_1 \gamma_\alpha
\gamma^5 \Gamma_2 iD_\lambda G_{\rho\sigma} h \nonumber\\
&& + \frac{4}{x^2} (v-\hat{q})^\beta \bar{h} \Gamma_1
\gamma^\alpha \Gamma_2 \gamma^\rho (iD_\alpha iD_\beta + iD_\beta
iD_\alpha) iD_\rho h \nonumber\\
&& + \frac{4}{x^2} \bar{h} \Gamma_1 (\xslash{v} -
\hat{\xslash{q}} + \hat{m}_s) \Gamma_2 \gamma^\rho (iD)^2 iD_\rho h \nonumber\\
&& - \frac{16}{x^3} (v-\hat{q})^\alpha (v-\hat{q})^\beta \bar{h}
\Gamma_1 (\xslash{v} -
\hat{\xslash{q}} + \hat{m}_s) \Gamma_2 \gamma^\rho iD_\alpha iD_\beta
iD_\rho h \nonumber\\
&& - \frac{2}{x^2} \hat{m}_s \bar{h} \Gamma_1 i
\sigma^{\alpha\beta} \Gamma_2 \gamma^\rho G_{\alpha \beta} iD_\rho h \nonumber\\
&& - \frac{2}{x^2} i \varepsilon^{\sigma\alpha\rho\beta}
(v-\hat{q})_\beta \bar{h} \Gamma_1 \gamma_\alpha
\gamma^5 \Gamma_2 \gamma^\lambda G_{\rho\sigma} iD_\lambda h 
\end{eqnarray}

\end{document}